# Influence of density dependence of symmetry energy on fragmentation

Karan Singh Vinayak, Mohinder Singh, Suneel Kumar[#]

*School of Physics and Material Science, Thapar University*
*Patiala*
[#]`Suneel.kumar@thapar.edu`

**Abstract**

The fragmentation of projectile and spectator is studied at the different incident energies using isospin dependent QMD model with reduced isospin dependent cross-section. Different systems have been used for the analysis of fragment production( IMF ). We have used enhanced constant isospin dependent cross-section to explain the experimental findings which is valid for soft equation of state. In addition to that we have tried to study the influence of density dependent symmetry energy on fragment production.

## I. Introduction

Nuclear Physics in general and heavy-ion collisions in particular is of central interest to carry out the investigation for the nature of equation of state. The study of multifragmentation gives us the insight into the reaction dynamics and the hot, dense nuclear matter formed during the reaction.

In general, the fragments produced during the collision depends upon the incident energy as well as on the impact parameter of the reaction. At low incident energies, reaction dynamics are dominated by the attractive nuclear mean field potential. With the increase in the incident energy, repulsive nucleon-nucleon scattering becomes important. As the frequent NN collision will result in the formation of free nucleons(FN's), light charged particle's(LCP's) and heavy mass fragments(HMF's). Here we study the formation of intermediate mass fragments (IMF's) at the incident energy of 600MeV/nucleon. The incident energy at which repulsive nucleon-nucleon scattering becomes dominant. As the composite mass dependence of the fragment production (IMF's) on the mean field can be sorted out by noticing the sensitivity of IMF production on system size and impact parameter. All the previous investigations reveal that the energy transfer in the IQMD events is governed by the elementary nucleon-nucleon cross-section, which was assumed to be same in the nuclear environment as in the free space. The exact nature of NN cross-section, on the other hand is still an open question [1]. A large number of calculations exist in the literature suggesting different strength and forms of the NN cross-sections. There has been a considerable progress during recent years in experimental studies. Therefore, we have reduced the in-medium cross-section to fit the theoretical predictions with experimental data.

In addition to that we have tried to study the fragment production by putting density constraint on symmetry energy. The symmetry energy tends to vary with the density as the reaction proceeds. The term symmetry energy $E(\rho)$ implies on an estimate of the energy cost to convert all the protons in a nuclear matter to the neutrons at a fixed density $\rho$ [2].

$$E(\rho) = E(\rho,1) - E(\rho,0)$$

Till now our understanding for the nucleon-nucleon interaction has come from studying the nuclear matter density at normal density ($\rho = 0.16$ fm$^{-3}$). As in the present scenario symmetry energy is taken as 32 MeV corresponding to normal nuclear matter density. It has been observed that nuclear matter density is below the normal density when fragmentation takes place. So, it is an interesting and important goal of heavy-ion physics to extract information about the symmetry energy of the nuclear matter at densities higher and lower than normal nuclear density. The equation below gives us the theoretical conjecture of how symmetry energy varies against $\rho$.

$$E(\rho) = E(\rho_o) \cdot (\rho/\rho_o)^{\gamma}$$

$\gamma$ tells us about the stiffness of the symmetry energy. For the present study different values of $\gamma$ have been taken for different systems at different incident energies. A comparable study have been performed to analyse the stiffness of the symmetry energy.

**IQMD Model :**

The Isospin-dependent Quantum Molecular Dynamic model is the refinement of QMD model based on event by event method. The reaction dynamics are governed by mean field, two-body collision and Pauli blocking.

The baryons are represented by Gaussian-shaped density distributions

$$f_i(\vec{r},\vec{p},t) = \frac{1}{\pi^2 h^2} \times e^{-[\vec{r}-\vec{r_i}(t)]^2 \frac{1}{2L}} \times e^{-[\vec{p}-\vec{P_i}(t)]^2 \frac{2L}{h^2}}$$

The successfully initialized nuclei are then boosted towards each other using Hamilton equations of motion.

$$\frac{dr_i}{dt} = \frac{d\langle H\rangle}{dp_i} \quad ; \quad \frac{dp_i}{dt} = -\frac{d\langle H\rangle}{dr_i}$$

With $\langle H\rangle = \langle T\rangle + \langle V\rangle$ is the total Hamiltonian.

$$\langle H\rangle = \sum_i \frac{p_i^2}{2m_i} + \sum_i\sum_{j>i} \int f_i(\vec{r},\vec{p},t) \times V^{ij}(\vec{r_i},\vec{r_j'})$$
$$\times f_j(r',p',t)\, d\vec{r}d\vec{r'}d\vec{p}d\vec{p'}$$

The total potential is the sum of the following specific elementary potentials.

$$V = V_{Sky} + V_{Yuk} + V_{Coul} + V_{mdi} + V_{loc}$$

During the propagation, two nucleons are supposed to suffer a binary collision if the distance between their centroid is

$$|r_i - r_j| \leq \sqrt{\sigma_{tot}/\pi}$$

Where $\sigma_{tot} = \sigma(\sqrt{s}, type)$

The collision is blocked with a possibility
$P_{block} = 1 - (1 - P_i) - (1- P_j)$
here $P_i$ and $P_j$ are the already occupied phase space fractions by other nucleons.

## Results and Discussion :

We here simulate the reactions of $_{54}Xe^{131} + _{79}Au^{197}$ and $_{57}La^{124} + _{50}Sn^{118}$ at the incident energies of 50 MeV/nucleon and 600 MeV/nucleon respectively.

At the incident energy of 50 MeV/nucleon, the nucleon mean field dominates the dynamics of the reaction. Indeed, at the incident energy of 600 MeV/nucleon repulsive nucleon-nucleon scattering dominates the dynamics of the reaction. The phase space generated by the IQMD model has been analysed by using the minimum spanning tree (MST)[3]. The MST method binds two nucleons in a fragment, if their distance is less than 4 fm. The entire calculations are performed at t = 200 fm/c. This time is chosen by keeping in view the saturation of collective flow.

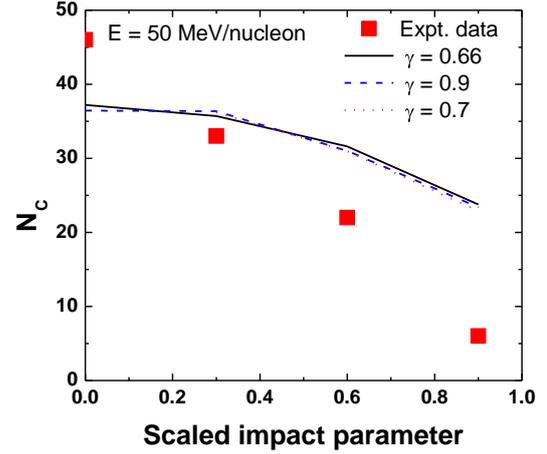

Fig. 1: Impact parameter dependence of the charged particle multiplicity at E/A = 50 MeV/nucleon for $_{54}Xe^{131}+_{79}Au^{197}$.

. In view of the findings from the Chen. et. al.[4], it is believed that the best estimate of the density dependence of the symmetry energy that can be extracted from the heavy-ion reaction studies is,
$E(\rho) = E(\rho_o) \cdot (\rho/\rho_o)^\gamma$ where $\gamma = 0.6 - 1.05$.

It can be interesting to carry out the investigation for the density dependence of symmetry energy for the different values of γ, i.e. the stiffness of symmetry energy. Therefore we have performed a comparable study by parameterizing the charge particle multiplicity against the scaled impact parameter for the various forms of the density dependence of the nuclear symmetry energy. In fig 1, we display the charge particle multiplicity $N_c$ as a function of scaled impact parameter for the reaction $_{54}Xe^{131} + _{79}Au^{197}$ at the incident energy of 50 MeV/nucleon for γ = 0.66, 0.9 and 0.7 respectively The trends observed through our simulations are somehow in accordance with the data. Although, the charged particle multiplicity $N_c$ seems to be less sensitive for the various parameterizations of symmetry energy i.e. for different values of γ.

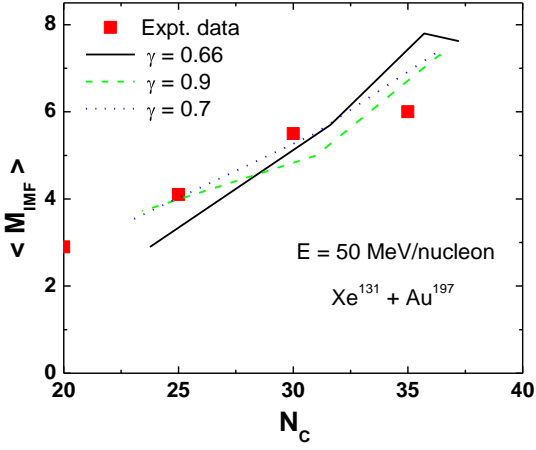 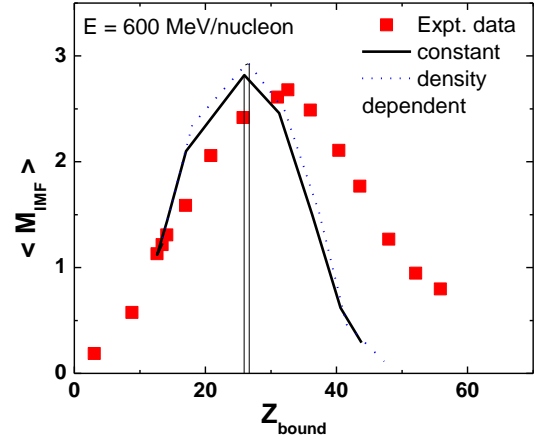

Fig. 2 : Correlation between the mean IMF multiplicity $M_{IMF}$ And charged particle multiplicity $N_c$ at E = 50 MeV/nucleon for the system $_{54}Xe^{131} + _{79}Au^{197}$.

In fig. 2, we display correlation between the mean IMF multiplicity $M_{IMF}$ and charged particle multiplicity $N_c$ at the incident energy of 50 MeV/nucleon for the similar system $Xe^{131} + Au^{197}$. It is interesting to note that by constraining the density dependence of symmetry energy for $\gamma = 0.7$, is best among the other two parameterizations in order to agree with the experimental data. This form of the density dependence of the symmetry energy is consistent with the parameterization adopted by Heiselberg and Hjorth-Jensen in their study on neutron stars [5].

Although we have tried to study the influence of density dependent symmetry on fragmentation in addition to reduced cross section in fig 3. The correlation of mean IMF multiplicityp with $Z_{bound}$ is plotted for the system $La^{124} + Sn^{118}$. For the present study $\gamma$ is taken 0.69 [6]. We have shown IMF's as a function of $Z_{bound}$. The quantity $Z_{bound}$ is defined as the sum of all atomic numbers ($Z_i$) of all projectile fragment with $Z_i \geq 2$. $Z_{bound}$ gives us a very good determination of the impact parameter. $Z_{bound}$ is a measure of the violence of the collision and of the energy deposited in the excited spectator.

The trends observed through our simulations are somehow in accordance with the data. Indeed, the maximum of IMF multiplicity is reached at semi peripheral collisions. The peak of maximum production is shifted slightly w.r.t data in case we consider symmetry energy as a function of the density. In case of central collisions at incident energy of 600 MeV/nucleon, the violent collision reduces the fragment production. While in case of the peripheral collisions the IMF production again decreases due to the lesser overlapping of the target and projectile. In such a case heavy mass fragments are produced. The peak value for IMF production is more as compared to the data. Indeed, the maximum of the IMF multiplicity is reached for significantly smaller values of $Z_{bound}$ i.e. at more central collisions**.**

Fig. 3: Correlation between the mean IMF multiplicity $M_{i.m.f}$ and $Z_{bound}$ for the systems $La^{124} + Sn^{118}$ at constant symmetry energy and density dependent symmetry energy for $\gamma = 0.69$ respectively

In future, we will try to study the influence of density dependent symmetry energy on fragmentation for the various systems having different mass and different isotopic compositions.

## Conclusion:-

In conclusion, we have investigated the effect of reduced cross section on the fragment production (IMF). The calculation with soft E.O.S. and reduced isospin dependent cross section result in the maximum IMF multiplicity at more central collisions. However, the peak value of the mean IMF multiplicity is underestimated about 10%. Our simulation with density dependent symmetry energy concludes a minor shift in peak IMF multiplicity.

### REFERENCES

[1] S.Kumar and R.K. Puri, Phys. Rev. C 58, 1628 (1998). S. Kumar, Rajni, and S. Kumar, Phys. Rev. C 82, 024610 (2010).
[2] Chan Xu, Bao-An Li and Lie-Wen Chen, Phys. Rev. C 82, 054607 (2010).
[3] S.Kumar, S.Kumar and R.K. Puri, Phys. Rev. C 81, 014601(2010); ibid Phys. Rev. C 81, 014611(2010).
[4] L. W. Chen, C. M. Ko. And B. A. Li, Phys. Rev. Lett. 94, 032701 (2005)
[5] H. Heiselberg and M. Hjorth-Jensen, Phys. Rep. 328, 237 (2000).
[6] D.V. Shetty, S.J. Yenello, G.A. Souliotis, Phys. Rev. C 76, 024606(2007).